\documentclass[12pt]{JHEP3}
\usepackage{graphicx}
\usepackage{amsmath}

\usepackage{braket}

\newcommand{\be}{\begin{equation}}
\newcommand{\ee}{\end{equation}}
\newcommand{\bea}{\begin{eqnarray}}

\newcommand{\eea}{\end{eqnarray}}
\newcommand{\ti}{\widetilde}

\newcommand{\gl}{\lambda}
\newcommand{\refe}[1]{Eqn.~(\ref{#1})}

\normalbaselines

\newcommand{\drawsquare}[2]{\hbox{%
\rule{#2pt}{#1pt}\hskip-#2pt
\rule{#1pt}{#2pt}\hskip-#1pt
\rule[#1pt]{#1pt}{#2pt}}\rule[#1pt]{#2pt}{#2pt}\hskip-#2pt
\rule{#2pt}{#1pt}}

\def\tv#1{\vrule height #1pt depth 5pt width 0pt}

\newcommand{\Yfund}{\raisebox{-.5pt}{\drawsquare{6.5}{0.4}}}
\newcommand{\Yasymm}{\raisebox{-3.5pt}{\drawsquare{6.5}{0.4}}\hskip-6.9pt%
        \raisebox{3pt}{\drawsquare{6.5}{0.4}}}

\title{\begin{center} 
General Gauge Mediation \\ and Deconstruction
\end{center}}
\author{
~~ Moritz McGarrie\\
Queen Mary University of London\\
Center for Research in String Theory\\
Department of Physics\\
Mile End Road, London, E1 4NS, UK.\\
\email{m.mcgarrie@qmul.ac.uk}
}

\abstract{We locate a supersymmetry breaking hidden sector and supersymmetric standard model on different lattice points of an orbifold moose.  The hidden sector is encoded in a set of current correlators and the effects of the current correlators are mediated by the lattice site gauge groups with ``lattice hopping'' functions and through the bifundamental matter that links the lattice sites together.  We show how the gaugino mass, scalar mass and Casimir energy of the lattice can be computed for a general set of current correlators and then give specific formulas when the hidden sector is specified to be a generalised messenger sector. The results reproduce the effect of five dimensional gauge mediation from a purely four dimensional construction.}

\preprint{QMUL-PH-10-12}

\begin{document}
\section{Introduction}
If in nature supersymmetry (SUSY) is broken in a hidden sector which is charged under the standard model gauge groups,\footnote{In a weakly coupled description of a hidden sector the charged fields are a messenger sector; in a strongly coupled hidden sector, these messengers may not be identifiable.} and that does not directly couple to standard model matter fields, then the hidden sector can be encoded in two point functions of currents which couple to the standard model gauge fields and generate an effective action \cite{Meade:2008wd,Buican:2008ws,deGouvea:1997tn,Intriligator:2008fr,Ooguri:2008ez,Distler:2008bt,Benini:2009mz,Buican:2009vv,Argurio:2009ge,Luo:2009kf,Kobayashi:2009rn,Lee:2010kb,Intriligator:2010be,Marques:2009yu,Antoniadis:2010nj,Argurio:2010fn}.   The effective action includes corrections to gauge boson and gaugino kinetic terms and corrections to the $D$ terms. The mediation of the kinetic terms of the effective action, in loops, will generate scalar masses of the standard model superpartners.  Additionally, some current correlators will generate gaugino mass terms.  As a result, gaugino masses are not a priori related to sfermion masses, resulting in six independent parameters and various mass sum rules \cite{Meade:2008wd,Carpenter:2010as}. 

The current correlators encode the hidden sector; changing the hidden sector model will change the current correlators and one may explore the parameter space of phenomenologically viable models \cite{Bhattacharyya:2010rm,Carpenter:2008he,Abel:2009ve,AbdusSalam:2009tr} and the mass ratio of gauginos and sfermions  \cite{Cheung:2007es,Buican:2008ws,Intriligator:2008fe, Marques:2008va,Komargodski:2009jf,Koschade:2009qu,Matos:2009xv,Dumitrescu:2010ha}.  A  different direction is to explore how the current correlators are mediated in the loop to the standard model. The standard four dimensional approach at leading order is to use the massless vector superfields. Semi-direct mediation makes the messenger sector an intermediate stage in the mediation, recursively embedding one set of current correlators inside another set of current correlators \cite{Seiberg:2008qj,Argurio:2009ge,Argurio:2010fn}. A variation of the original direct mediation is to mediate the effects of the current correlators by massive vector superfields generated via a super Higgs mechanism \cite{Gorbatov:2008qa,Intriligator:2008fr,Buican:2009vv,Intriligator:2010be,Green:2010ww}, such as may arise in Grand Unified Theories.  In more complete models, both massless and Higgsed mediators may need to be combined \cite{Matos:2010ie}.  Another approach is to locate the hidden sector and visible standard model matter on different ends of an extra dimensional interval $S^1/\mathbb{Z}_{2}$, the Mirabelli-Peskin model \cite{Mirabelli:1997aj,McGarrie:2010kh}, with bulk gauge fields, in which case the full Kaluza-Klein (KK) tower of vector superfields may mediate the effects. Additionally, one may warp the background metric, for instance a slice of $AdS_{5}$ \cite{McGarrie:2010yk}. The general effect of this is to introduce some additional mass scale, say the vacuum expectation value of the super Higgs or the length of the interval\footnote{Compactifying $5D$, $\mathcal{N}=1$ super Yang-Mills on an orbifold generates a $4D$ $\mathcal{N}=1$ vector superfield and a negative parity chiral superfield. The masses of the K.K. tower of vector superfields are generated by ``eating'' the chiral superfield, so these models are not unrelated.} which suppress the momentum in the loop generating sfermion masses and therefore suppress sfermion masses.  We want to stress that for any given hidden sector one uses the same set of current correlators, regardless of the variation of the mediation type listed above.  The suppression arises when the current correlators are expanded in the limit of $\frac{p^2}{M^2}\rightarrow 0$, where $M$ is a characteristic scale of the hidden sector \cite{McGarrie:2010kh}.

In this paper we reconsider lattice (de)constructions \cite{Hill:2000mu,ArkaniHamed:2001ca,ArkaniHamed:2001nc,Cheng:2001vd,Cheng:2001nh,Kunszt:2004ps,DiNapoli:2006kc} that generate an orbifold \cite{Csaki:2001em}. The aim of this paper is to combine the framework of ``General gauge mediation'' \cite{Meade:2008wd} with the supersymmetric orbifold lattice construction \cite{Csaki:2001em} and recover a lattice description of ``General gauge mediation in five dimensions'' \cite{McGarrie:2010kh}. We locate the hidden sector and visible standard model on opposite lattice end points of a quiver gauge diagram with $N$ lattice sites.  Each lattice site is a copy of a four dimensional supersymmetric standard model parent gauge group $SU(5)$. Each lattice site is connected using bifundamental chiral superfields that link the lattice together.  The combination of super Yang-Mills with specified ``lattice hopping" vectors and the bifundamental linking matter, will mediate the supersymmetry breaking effects from the hidden sector lattice site to the standard model lattice site, to generate sfermion masses \cite{Csaki:2001em,Cheng:2001an}.  In this model, the lattice site spacing $a$ will be the scale that suppresses the momentum in the loops for sfermion masses. We shall see that effective five dimensional suppression effects arise when $\frac{1}{(Na)^2}\ll M^2 $ and that we recover four dimensional effects when $M^2 \ll \frac{1}{(Na)^2}$. At low energies this lattice construction is equivalent to the five dimensional orbifold model mentioned above.  The scale $a$  arises in the masses of the propagating gauge fields and bifundamental matter which resemble KK states at low energies.  The masses of the KK spectrum are generated by the vacuum expectation value of the scalars of the bifundamental chiral superfields that link the lattice together and so we see that these suppression methods are all related.

This paper is organised as follows: in Section \ref{section:Framework} we layout the framework of the orbifold lattice, including the lattice propagators and the mass spectrum.  Section \ref{section:currents} locates an arbitrary hidden sector on a lattice site, generates the current correlators and then formulas for the gaugino masses, sfermion masses and Casimir energy of the lattice.  In section \ref{section:general} we give a concrete description of a hidden sector in terms of generalised messengers coupled to a supersymmetry breaking spurion.  Specifying the hidden sector allows the current correlators to be evaluated and we produce the gaugino masses, sfermion masses and Casimir energy of the lattice for this specific hidden sector.  In \ref{conclusion} we summarise and conclude our discussion.
\section{Framework}\label{section:Framework}
This section concisely reviews the construction of the orbifold moose (quiver gauge diagram) following the work of \cite{Csaki:2001em}. We start with a lattice of four dimensional, $SU(5)_{i}$ super Yang-Mills gauge groups all identified with the gauge groups of the supersymmetric standard model.  The matter content of the lattice is:
\begin{equation}
 \begin{array}{c|cccccccc}
        & SU(5)_0 & SU(5)_1 & \cdots & SU(5)_{N-2} & SU(5)_{N-1} \\ \hline
\tv{15} \tilde{P}_1, \ldots, \tilde{P}_5
        & \overline{\Yfund} & 1 & \cdots & 1 & 1 \\
           Q_1 & \Yfund  & \overline{\Yfund} & \cdots & 1 & 1 \\
        \vdots & \vdots & \vdots & \ddots & \vdots & \vdots \\
         Q_{N-1} & 1 & 1 & \cdots & \Yfund & \overline{\Yfund} \\
P_1, \ldots, P_5 & 1 & 1 & \cdots & 1 & \Yfund \\
\overline{\bf 5}_{1,2,3} & 1 & 1 & \cdots & 1 & \overline{\Yfund} \\
{\bf 10}_{1,2,3} & 1 & 1 & \cdots & 1 & \Yasymm \\
  H_d & 1 & 1 & \cdots & 1 & \overline{\Yfund} \\
  H_u & 1 & 1 & \cdots & 1 & \Yfund \\
\Phi & \Yfund & 1 & \cdots & 1 &  1\\ 
 \tilde{\Phi} & \overline{\Yfund} & 1 & \cdots & 1 &  1\\
  X & 1 & 1 & \cdots & 1 & 1 \\
 \end{array} \nonumber 
\end{equation}
The $Q_{i \alpha}^{\phantom{\beta}\beta}$ are bifundamental chiral superfields that link the $SU(M)^N$ lattice sites together. The $\alpha,\beta$ are gauge indices labelling the fundamental or antifundamental of the gauge group $SU(M)$.  The bifundamental scalars all obtain a vacuum expectation value $v$, which may be generated by some dynamical superpotential \cite{ArkaniHamed:2001ca,Csaki:2001em}.  The indices $i,j,k$ label the lattice, running from $i=0$ to $N-1$ (mod $N$) with lattice spacing $a=1/(\sqrt{2}g v)$, such that $\ell=N a$ is the length of the lattice. The five chiral multiplets $\tilde{P}$ and five $P$ are localised on the endpoint lattice sites and are required to cancel anomalies due to the breaking of ``hopping" symmetry at the lattice end points.  The $5,10$ and $H_{u},H_{d}$ play the role of the standard model matter and Higgs superfields. Additionally, we may add the fields $\Phi,\tilde{\Phi}$ as messenger superfields coupled to a spurion $X=\braket{X}+\theta^2 F$. These fields will enter the discussion when we specify a generalised hidden sector in a later section. 

The bifundamental scalars are given a vacuum expectation value and fluctuations about that value, $Q_{i \alpha}^{\phantom{\beta}\beta}=v \delta_{\alpha}^{\phantom{\beta}\beta}+ \phi_{i \alpha}^{\phantom{\beta}\beta}$. The scalar kinetic terms of the bifundamental matter are used to generate a mass matrix for the gauge bosons, via the Higgs mechanism. The mass spectrum is computed in \cite{Csaki:2001em} and in \cite{ArkaniHamed:2001ca,Hill:2000mu,Cheng:2001vd}.   For the gauge bosons, the masses are 
\be
m^2_{k}=8g^2 v^2 \sin^2 (\frac{k\pi}{2N}) \phantom{AAAAA} k=0, ..., N-1.
\ee
A key attribute of this setup is that lattice eigenstates are not mass eigenstates of the system.  
The mass eigenstates are given by 
\be
\tilde{A}_{k}=\sqrt{\frac{2}{2^{\delta_{k0}}N}}\sum_{j=0}^{N-1} \cos \frac{(2j+1)k\pi}{2N}A_{j} \phantom{AAAAA} k=0, ..., N-1. 
\ee
These are even parity modes.    The fermion mass matrix of bifundamental fermions $q_{i}$ and gauginos $\gl_{i}$ must be diagonalised as in \cite{Csaki:2001em}. The even states  $\gl_{i}$ have masses
\be
 m^2_{k}=8g^2 v^2 \sin^2 (\frac{k\pi}{2N}) \phantom{AAAAA} k=0, ..., N-1
\ee
and eigenvectors
\be
\tilde{\gl}^{+}_{k}=\sqrt{\frac{2}{2^{\delta_{k0}}N}}\sum_{j=0}^{N-1} \cos \frac{(2j+1)k\pi}{2N} \gl_{j} \phantom{AAAAA} k=0, ..., N-1. 
\ee 
The odd parity fermions $q_{i}$ have masses
\be
 m^2_{k}=8g^2 v^2 \sin^2 (\frac{k\pi}{2N}) \phantom{AAAAA} k=1, ..., N-1
\ee
with eigenstates 
\be
\tilde{\gl}^{-}_{k}=\sqrt{\frac{2}{N}}\sum_{j=1}^{N-1} \sin \frac{j k\pi}{2N} q_{j} \phantom{AAAAA} k=1, ..., N-1. 
\ee
In the continuum limit, the fermions $q_{i}$ should coincide with the adjoint fermion $\chi$  of a negative parity chiral superfield.  To construct this adjoint fermion one identifies the bifundamental scalar with the unitary link variable: $Q_{\alpha i}^{\phantom{\beta}\beta}=v U_{ \alpha i}^{\phantom{\beta}\beta}$ where $U$ is a unitary matrix. Keeping track of indices  we may relabel $(U_{\alpha}^{\phantom{\beta} \beta})_{i}^{\dagger}q_{\gamma}^{\phantom{\beta}\beta}=\chi_{\gamma i}^{\phantom{\beta}\alpha}$, where both indices $\alpha,\gamma$ are valued at the $i$'th lattice site. A negative parity adjoint scalar $\Sigma$ is similarly defined and as the multiplet is supersymmetric, the mass spectrum and eigenstates are equivalent to that of the fermion.  

Using the above eigenfunctions, the mixed space scalar propagator can readily be determined by insertion of a closure relation for the mass eigenstates.   The result is
\be
\braket{p^2;k,l} =\frac{2}{N}\sum_{j=0}^{N-1}\frac{1}{2^{\delta_{j0}}}\cos (\frac{(2k+1)j\pi}{2N})\cos (\frac{(2l+1)j\pi}{2N})\frac{1}{p^2+(\frac{2}{a})^2\sin^2 (\frac{j\pi}{2N})}.\label{propagator1}
\ee
In summary, the resulting low energy degrees of freedom and field content for large $N$ is that of an $\mathcal{N}=1$ positive parity vector multiplet and negative parity chiral superfield of $\mathcal{N}=1$ super Yang-Mills in five dimensions compactified on $R^{1,3}\times S^1/\mathbb{Z}_{2}$ \cite{Hebecker:2001ke,McGarrie:2010kh}.

\subsection{The periodic lattice}
It is useful to compare this construction with that of a periodic lattice corresponding to a $5d$ theory compactified on a circle \cite{Csaki:2001em}. 
The anomaly cancelling $P$ and $\tilde{P}$ fields are unwanted in the periodic construction. All the bulk fields have a mass spectrum given by
\be
m^2_{k}=8g^2 v^2 \sin^2 (\frac{k\pi}{N}) \phantom{AAAAA} k=0, ..., N-1.
\ee
The vector superfield mass eigenstates are also related to the lattice eigenstates through
\be
\tilde{V}_{k}=\frac{1}{\sqrt{N}}\sum_{j=0}^{N-1} e^{i(2\pi k j)/N}V_{j} \phantom{AAAAA} k=0, ..., N-1.
\ee
The mixed space scalar propagator for the circle may also be written as
\be
\braket{p^2;k,l} =\frac{1}{N}\sum_{j=0}^{N-1}e^{-i(2\pi k j)/N}e^{i(2\pi \ell j)/N}\frac{1}{p^2+m^2_{j}}.\label{propagator2}
\ee
The low energy degrees of freedom are of a full $4d$ $\mathcal{N}=2$ model, including restoring the zero modes that had been projected out by negative parity, in the interval case. 
\section{Lattice localised currents} \label{section:currents}
This section will encode a SUSY breaking sector, localised on the lattice site i=0 in terms of current correlators \cite{Meade:2008wd}.  The lattice has a set of vector superfields $\{V_{i}\}$ and a set of current multiplets $\{\mathcal{J}_{i}\}$ associated to the bifundamental matter linking the lattice\footnote{In this discussion we will ignore the currents of the fields $P_{i}$ and $\tilde{P}_{i}$.}.  The supersymmetric standard model matter (visible sector) will form a current multiplet $\mathcal{J}^v_{N-1}$. Additionally, we may locate a hidden sector at lattice point $i=0$, $\mathcal{J}^{h}_{0}$. 

We may couple the hidden sector current multiplet to the lattice gauge fields
\be 
S_{int}=2g\!\int\! d^4 x  d^{4}\theta \mathcal{J}^h_{0}
\mathcal{V}_{0}= g \int d^4 x (JD_{0}- \gl_{0} j \!-
 \!\bar{\gl}_{0} \bar{j}-j_{\mu}A^{\mu}_{0})\ee
As this whole discussion refers to the hidden sector current multiplet $\mathcal{J}^{h}_{0}$ at $i=0$, we will drop the hidden sector index.  The change of the effective Lagrangian to
$O(g^{2})$ is
\begin{align}  
\delta \mathcal{L}_{eff}= &-g^{2} \tilde{C}_{1/2}(0) i \lambda_{0} \sigma^{\mu} \partial_{\mu} \bar{\lambda}_{0}
- g^{2}\frac {1} {4} \tilde{C}_1(0) F_{0 \mu\nu} F^{\mu\nu}_{0} \nonumber + \frac{1}{2}g^2 \tilde{C}_{0}(0)D^2_{0}\\
&-g^{2}\frac {1}{ 2}(M \tilde{B}_{1/2}(0) \lambda_{0} \lambda_{0} + M \tilde{B}_{1/2}(0)\bar{\gl}_{0}\bar{\gl}_{0}).\nonumber
\end{align}
The $\tilde{B}$ and $\tilde{C}$ functions are related to momentum space current correlators \cite{Meade:2008wd,McGarrie:2010kh} and are located on the i=0 lattice site: 
\begin{align}
\braket{J(p)J(-p)} =&\tilde{C}_0(p^2/M^2;M/\Lambda)  \label{eq:c0}   \\
\braket{j_\alpha(p)\bar j_{\dot\alpha}(-p) } =&-\sigma_{\alpha\dot\alpha}^\mu p_\mu\tilde{C}_{1/2}(p^2/M^2;M/\Lambda) \label{eq:c1/2}		\\ 
\braket{j_\mu(p)j_\nu(-p)} =&-(p^2\eta_{\mu\nu}-p_\mu p_\nu)\tilde{C}_1(p^2/M^2;M/\Lambda)	\label{eq:c1}	\\ 
\braket{j_{\alpha}(p)j_{\beta} (-p)} =&\epsilon_{\alpha\beta}M\tilde{B}_{1/2}(p^2/M^2)     \label{eq:b1/2}	
\end{align}
$M$ is some characteristic mass scale of the theory. $\tilde{C}_s$ and $\tilde B$ are Fourier transforms of $C_s$ and $B$:
\begin{equation}
\begin{split}
\tilde{C}_s\left(\frac{p^2}{M^2};\frac{M}{\Lambda}\right)&=\int d^4 x e^{ipx} \frac 1 {x^4} C_s(x^2 M^2)\\
M\tilde{B}_{1/2}\left(\frac{p^2}{M^2}\right)&=\int d^4 x e^{ipx} \frac 1 {x^5} B_{1/2}(x^2 M^2)\;.
\end{split}
\end{equation}  $\Lambda$ is a UV cutoff regulating the integrals. $C_s$ and $B$ are the position space current correlators 
\begin{align}
\braket{J(x)J(0)}=&\frac{1}{x^{4}}C_0(x^{2}M^{2}) \\
\braket{j_\alpha(x)\bar{j}_{\dot\alpha}(0)}=&-i\sigma_{\alpha\dot\alpha}^\mu \partial_\mu(\frac{1}{x^{4}}C_{1/2}(x^{2}M^{2}))\\
\braket{j_\mu(x)j_\nu(0)}=&(\partial^2\eta_{\mu\nu}-\partial_\mu \partial_\nu)(\frac{1}{x^{4}}C_1(x^{2}M^{2}))\\
\braket{j_\alpha(x)j_\beta(0)}=&\epsilon_{\alpha\beta}\frac{1}{x^{5}}B_{1/2}(x^{2}M^{2}) .\label{scale}
\end{align}
If supersymmetry is unbroken 
\begin{equation}
C_0=C_{1/2}=C_1\;,\qquad \text{and} \qquad B_{1/2}=0\;
\end{equation}
and in this case the $C_{s}$ terms may be related to the change in the beta function of $g$ due to the hidden sector matter at the lattice site $i=0$.  Key to this whole discussion is that whilst the hidden sector may be strongly coupled in which case we must determine the correlators exactly, we may still work perturbatively in the coupling $g$ of the lattice to mediate the effects of these current correlators to the visible sector.  If the hidden sector has a weakly coupled description, we may additionally determine the structure of the correlators perturbatively, which we demonstrate in section \ref{section:general}.  In the next subsections we will demonstrate how these current correlators may be used to determine soft mass formulas and the Casimir energy of the lattice.
\subsection{Gaugino masses}
A soft supersymmetry breaking gaugino mass for the $i=0$ lattice sites arises at tree level. It is given by 
\be 
m_{\gl, i=0}= g^2 M\tilde{B}_{1/2}(0) \label{gauginomass1}
\ee
In the continuum orbifold scenario \cite{McGarrie:2010kh}, the orbifold fixed points break Lorentz invariance in the fifth dimension and this is signified by the ``brane''\footnote{The word ``brane'' being used to denote the ends of an interval, where matter is located in five dimensional orbifold constructions.} localised current correlators not preserving incoming and outgoing $p_{5}$ momenta.  These current correlators therefore couple, equally, to all states of the K.K. tower of vector superfields.  In the lattice picture the currents only coupled to fields at a single lattice site.  However it is precisely because the lattice fields are a sum of mass eigenstates, that the current correlator still generates a correction to all mass eigenstates in the lattice picture. The soft term mass must be included in the full mass matrix of all fermions ($\gl_{i}, q_{i}$) in the lattice and if we assume that $m_{\gl, i=0}$ is small, we may treat this as a perturbation of the full mass matrix.  This process is outlined in \cite{Csaki:2001em} and one finds the zero mode mass is
\be
 m_{0}= \frac{g^2}{N} M\tilde{B}_{1/2}(0).
\ee
The four dimensional coupling is determined from $g^2_{4d}=g^2/N$.  The mass splittings are also similarly obtained
\be
m^2_{k} =4 g^2 v^2 (\sin \frac{k\pi}{2N}\pm \frac{g}{\sqrt{2}vN}M\tilde{B}_{1/2}(0)\cos^2 \frac{k\pi}{2N})\sin\frac{k\pi}{2N}, \phantom{AAAA} k=1,...,N-1.
\ee
Heuristically, we see that it is the process of moving from lattice states to mass eigenstates that reproduces the orbifold effect of a soft mass coupling to all K.K. modes.
\subsection{Sfermion masses}
\FIGURE{
\centering
\includegraphics[scale=0.8]{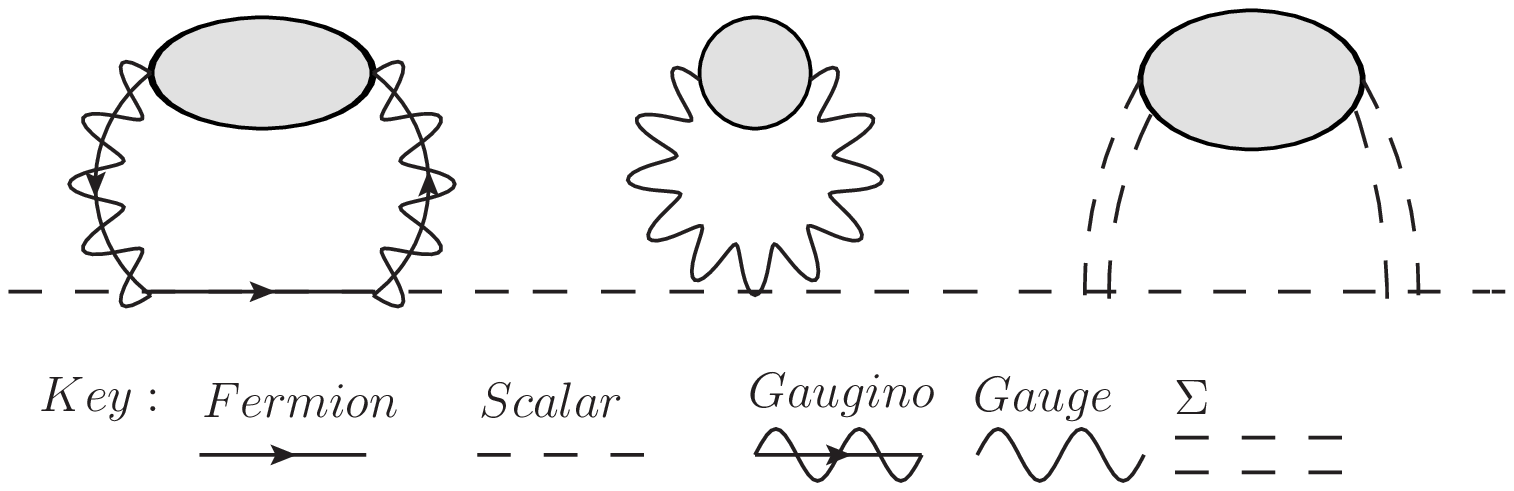}
\caption{The graphical description of the contributions of the two point functions
to the sfermion soft masses.  The ``blobs'' represent hidden sector current correlators on the lattice site $i=0$.  From left to right the diagrams represent the correlator $\braket{j_\alpha \bar j_{\dot\alpha}}$ mediated by the ``lattice hopping" gauge bosons, $\braket{j_\mu j_\nu }$ by the gaugino and  $\braket{JJ}$ by the adjoint scalar built from the bifundamental linking scalars fields.  The scalar and fermion lines at the bottom are located at the lattice site $i=N-1$.}
\label{sfermion3}
}
We would like to propagate the effects of supersymmetry breaking from the $i=0$ lattice site to the $i=N-1$ site to generate scalar masses for the MSSM located at that site.  This is a loop diagram with intermediate gauge boson, gaugino and $D_{i}$ term as can be seen in figure \ref{sfermion3}.   The gauge boson and gaugino are dynamical and they may propagate around the lattice using the ``lattice hopping'' wavefunction.  The lattice scalar propagator \refe{propagator1} from the hidden sector lattice site to visible lattice site is 
\be 
\braket{q; 0,N-1}=\frac{2}{N}\sum_{j=0}^{N-1}\frac{1}{2^{\delta_{j0}}}\frac{(-1)^{j}\cos^{2}{\frac{j\pi}{2N}}}{q^{2}+(\frac{2}{a})^{2}{\sin^2{\frac{j\pi}{2N}}}} =a^2\prod_{j=0}^{N-1}\frac{1}{(aq)^{2}+4\sin^{2}{\frac{j\pi}{2N}}} .
\ee 
For the periodic lattic case the same final equation holds, after adjusting for the mass eigenstates by $2N\rightarrow N$ in the denominator.  We highlight that from interval to interval fixed point propagators, the factor $(-1)^{j} $ arises.  This factor is crucial in the cancellation of alternate states of the K.K. tower which generates the suppression of sfermion masses at large momenta and keeps ``brane to brane'' diagrams UV finite.  This factor does not arise in ``brane'' to same ``brane'' diagrams, which is why the vacuum energy diagram is divergent.  Equally, diagrams with a double insertion of the gaugino mass will also be UV divergent.  To generate a gaugino or gauge boson propagator we supplement this scalar propagator with the correct Lorentz structure of a fermion, $\sigma_{\alpha\dot\alpha}^\mu \partial_\mu$ or transverse gauge projector $P^{\mu\nu}= (\partial^2\eta^{\mu\nu}-\partial^\mu \partial^\nu)/\partial^2$ as necessary.  The remaining diagram propagated by $\Sigma$, is not so simple. In orbifold models \cite{McGarrie:2010kh}, the $D$ term is given by $D= (\partial_{5}\Sigma+iX^3)$, where $\Sigma$ is a dynamical negative parity scalar field and $\partial_{5}$ is a derivative in the fifth direction. This field contributes to the propagation of supersymmetry across the interval.  To make the $D_{i=0}$ dynamical on the lattice, one may integrate out this auxiliary field in terms of the lattice currents and find 
\be 
D_{i}=gJ_{i}= g \text{tr}((Q_{\alpha}^{\beta})^{\dagger}_{i}T^{a \gamma}_{\alpha} Q_{\gamma i}^{\phantom{\beta}\beta}-Q^{\phantom{\beta}\beta}_{\alpha i-1}T_{\beta}^{a\gamma} (Q_{\alpha}^{\phantom{\gamma}\gamma })^{\dagger}_{i-1}).
\ee
The index $a$ is a generator index running from $1$ to $N^2-1$.  Next, using $Q_{\alpha i}^{\phantom{\beta}\beta}=v U_{ \alpha i}^{\phantom{\beta}\beta}$ where $U$ is a unitary matrix and keeping track of the indices of each gauge group one finds 
\be 
D_{i}= \frac{1}{\sqrt{2}a}(T^a \Sigma_{i}-\Sigma_{i-1} T^a )+...
\ee
where we have relabelled $(U_{\alpha}^{\phantom{\beta} \beta})_{i}^{\dagger}Q_{\gamma}^{\phantom{\beta}\beta}=\Sigma_{\gamma i}^{\phantom{\beta}\alpha}$.  The contraction of indices has resulted in an adjoint scalar $\Sigma$ of the $i$'th lattice site and both indices $\alpha,\gamma$ are valued at the same lattice site. In the continuum limit, the $D$ term is a lattice derivative  of $\Sigma$ \cite{Bauer:2003mh}:
\be
\frac{1}{\sqrt{2}}\partial_{5}\Sigma = \frac{1}{\sqrt{2}} \lim_{a\rightarrow 0}\frac{\Sigma(y+a)-\Sigma(y)}{a}.
\ee
We may now associate $\Sigma$ with the negative parity scalar of the $5D$ $\mathcal{N}=1$ super Yang-Mills action.  An additional manipulation used to calculate this diagram in orbifold models, is to use 
\be 
\delta(0)= \frac{1}{2\ell} \sum_{n}\frac{p^2-m^2_{n}}{p^2-m^2_{n}}
\ee
to exchange $m^2_{n}$ terms, generated by the derivative $\partial_{5}$, for $p^2$. Whilst this manipulation is rather less precise on the lattice than in the continuum limit, it is \emph{necessary} to ensure that corrections to the beta function due to a supersymmetric hidden sector cannot generate sfermion masses. Collecting the contributions from all three diagrams, the sfermion formula is 
\begin{equation}
m_{\tilde{f}}^2= \sum _r g_{r}^4 c_2(f;r)E_r
\end{equation}
where
\begin{equation}
E_r\!\!=\!\! -\! \!\int\!\! \frac{d^4p}{ (2\pi)^4}p^{2}\!\braket{p^2;0,N\!-\!1}\! \!\braket{p^2;0,N\!-\!1}\!
[3\tilde{C}_1^{(r)}(p^2\!/M^2)-4\tilde{C}_{1/2}^{(r)}(p^2\!/M^2)+\tilde{C}_{0}^{(r)}(p^2\!/M^2)]. 
\end{equation}
For the interval model, using \refe{propagator1} gives 
\begin{equation}
E_r\!\!=\!\! -\! \!\int\!\! \frac{d^4p}{ (2\pi)^4}a^4p^{2}\!\left[\prod_{j=0}^{N-1}\!\frac{1}{(ap)^{2}+4\sin^2{\frac{j\pi}{2N}}}\right]^2\!\!\!\!
[3\tilde{C}_1^{(r)}(p^2\!/M^2)-4\tilde{C}_{1/2}^{(r)}(p^2\!/M^2)+\tilde{C}_{0}^{(r)}(p^2\!/M^2)]. \label{sfermionmasses}
\end{equation}
The same equation holds in the periodic case, using the periodic mass eigenstates $4\sin^2{\frac{j\pi}{N}}$. In this equation, we have used a standard model lattice with gauge coupling $g_{r}$ where $r=1,2,3$ refering to the group $U(1),SU(2),SU(3)$ respectively. 
$c_2(f;r)$ is the quadratic Casimir of the representation $f$ of the scalar mass in question, under the gauge group $r$. The integral is UV and IR finite. As discussed in the introduction, one can see that the momentum integral in this equation will be suppressed by the product of KK propagators with length scale $a$ entering from the mass of the KK modes.  The limit in which there is a single lattice site ($N=1$) is the corresponding four dimensional limit.  We find 
\begin{equation}
E_r= -\! \!\int\! \frac{d^4p}{ (2\pi)^4}\frac{1}{p^2}
[3\tilde{C}_1^{(r)}(p^2/M^2)-4\tilde{C}_{1/2}^{(r)}(p^2/M^2)+\tilde{C}_{0}^{(r)}(p^2/M^2)]. \label{4dlimit}
\end{equation}
This equation reproduces exactly the result of ``General gauge mediation'' \cite{Meade:2008wd}.  For $N=2$ lattice sites, we may take the two gauge eigenmasses to be $m_{0}=0$ and $m_{1}=m_{v}$.  For the sfermion mass formula, one obtains\footnote{We hope this result clarifies the connection between this paper and the more recent papers \cite{Auzzi:2010mb,Sudano:2010vt,Auzzi:2010xc}.}
\begin{equation}
E_r= -\! \!\int\! \frac{d^4p}{ (2\pi)^4}\frac{1}{4p^2}[\frac{m^2_{v}}{p^2+m^2_{v}}]^2
[3\tilde{C}_1^{(r)}(p^2/M^2)-4\tilde{C}_{1/2}^{(r)}(p^2/M^2)+\tilde{C}_{0}^{(r)}(p^2/M^2)]\label{n=2limit}
\end{equation}
where the unwanted factor of $1/4$ is absorbed into the gauge coupling $g^2_{4d}=g^2/N$.
\subsection{The Casimir energy}
In a globally supersymmetric theory the vacuum energy is zero. Supersymmetry breaking effects of the hidden sector will generate a vacuum energy and the lattice dependent part of this vacuum energy will correspond, in the continuum limit, to the Casimir energy of a higher dimensional theory \cite{Bauer:2003mh,Mirabelli:1997aj,McGarrie:2010kh}.  To calculate the Casimir energy we must compute the vacuum diagrams that appear in figure 2 of \cite{Intriligator:2008fr}. Each vacuum diagram may be generated by simply forming a closed loop with the field that propagates each current correlator in the effective action. The propagation is from the zeroth lattice site back to the zeroth lattice site in the loop. The zero to zero lattice site propagator is given by
\be
 \braket{q^2; 0,0}=\frac{1}{N q^2}  [1+\sum_{k=1}^{N-1} \frac{2(aq)^2 \cos^2 \frac{k\pi}{2N} }{(aq)^2+4\sin^2 \frac{k \pi}{2N} }].
\ee
This time there is no product form for the propagator as the $(-1)^j$ is absent.  The vacuum energy is given by
\be 
\frac{E_{vac}}{V_{4d}}\! =\!\sum_{r}g^2_{r}d_{g}\!\!\int\!\! \frac{d^4p}{ (2\pi)^4} p^2\braket{p^2; 0,0}
[3\tilde{C}_1^{(r)}(\frac{p^2}{M^2})-4\tilde{C}_{1/2}^{(r)}(\frac{p^2}{M^2})+\tilde{C}_{0}^{(r)}(\frac{p^2}{M^2})]. \label{VAC}
\ee
$d_{g}$ is the dimension of the adjoint representation of the gauge group $r$. This integral is UV divergent. To extract from this the finite Casimir energy, one must extract the continuum limit of this sum. The prescription for this is found in \cite{Bauer:2003mh}.  Additionally, the appendix includes relevant steps which are applied in the next section, where we focus on a generalised messenger sector for which the $\tilde{C}_{s}$ terms may be determined.

\section{Generalised messenger sector}\label{section:general}
In this section we give a concrete description of matter content of the SUSY breaking sector located at the zeroth lattice site, following the construction of \cite{Martin:1996zb}. We consider sets of $N$ chiral superfield messengers\footnote{We hope that this index $i$ which runs $1$ to $N$, the number of messengers, does not get confused with the lattice site index of the previous section.}  $\Phi_{i},\tilde{\Phi}_{i}$ in the vector like representation of the lattice gauge group, coupled to a SUSY breaking spurion $X= M + \theta^2 F $ with $F \ll M^2$. Generalisations to arbitrary hidden sectors are a straightforward application of the results of \cite{Marques:2009yu,McGarrie:2010kh}.  The superpotential is 
\be W  =X \eta_{i}\
\Phi_i \ti \Phi_i \label{superpotential2}\ee
In principle $\eta_{ij}$ is a generic matrix which may be diagonalised to its eigenvalues $\eta_{i}$ \cite{Martin:1996zb}.  The messengers will couple to the bulk vector superfield as
\be \delta {\cal L} = \int d^2\theta d^2\bar\theta \left(\Phi^\dag_i
e^{2 g V^a T^a} \Phi_i + \ti\Phi^\dag_i e^{-2 g V^a T^a}
\ti\Phi_i\right) + \left(\int d^2\theta\  W +
c.c.\right) \label{hiddensector} \ee
We can extract the multiplet of currents from the kinetic terms in the above Lagrangian. The current correlators can then be computed and their results can be found in \cite{Meade:2008wd,McGarrie:2010kh,Marques:2009yu}.  We will use the result of these current correlators to determine the gaugino massses, sfermion masses and Casimir energy.
\subsection{Gaugino masses}
The SUSY breaking zero mode Majorana gaugino mass is found by first evaluating the current correlator in \refe{gauginomass1}, which may be found in \cite{Martin:1996zb,Meade:2008wd,McGarrie:2010kh}, and then diagonalising the full fermion mass matrix and extracting the zero mode as discussed in the previous section.  For the zero mode this simply fixes $g^2/N=g^2_{4d}$.  The zero mode gaugino mass is found to be 
\be 
m^r_{\gl_{0} }= \frac{\alpha_{r}}{4\pi} \Lambda_{G}\ , \ \ \ \  \Lambda_{G}=\sum_{i=1}^{N}[\frac{d_{r}(i) F}{M}g(x_{i})]
\ee
The label $r=1,2,3$ refers to the gauge groups $U(1),SU(2),SU(3)$, $d_{r}(i)$ is the Dynkin index of the representation of $\Phi_{i},\tilde{\Phi}_{i}$ and
\be 
g(x)= \frac{(1-x)\log(1-x)+(1+x)\log(1+x)}{ x^2}
 \ee
where  $x_{i}=\frac{F}{\eta_{i}M^2}$.  $g(x)\sim 1$ for small $x$ \cite{Martin:1996zb}.

\subsection{Sfermion masses}
The entirely four dimensional limit of the sfermion mass formula when both the hidden and visible sector are located on the same single lattice site, is displayed in \refe{4dlimit}.  For the generalised messenger sector, this four dimensional result can be found in \cite{Martin:1996zb}. To obtain an effective five dimensional behaviour from the lattice, one must require sufficient lattice sites to suppress large contributions to loop momenta in the diagrams contributing to sfermion masses.  We start with \refe{sfermionmasses} and when $\frac{1}{(Na)^2}\ll M^2 $, one may then expand the current correlators in the limit $\frac{p^2}{M^2} \rightarrow 0$ and find \cite{McGarrie:2010kh}
\be
 [3\tilde{C}_1(p^2/M^2)-4\tilde{C}_{1/2}(p^2/M^2)+\tilde{C}_0(p^2/M^2)]\approx -\frac{1}{(4\pi)^2}\frac{2 d}{3}x^2 h(x)+O(p^2) \label{cterms}
\ee  
which is independent of $p^2$ at this order.  The function $h(x)$ is given by
\be h(x) =\frac{3}{2}[\frac{4+x-2x^2}{x^4}\log(1+x)+\frac{1}{x^2}] + (x\rightarrow -x),
\ee
where $h(x)$ for $x<0.8$ can be reasonably approximated by $h(x)=1$.  We find
\begin{equation}
m_{\tilde{f}}^2= \sum _r g_{r}^4 c_2(f;r)E_r
\end{equation}
where 
\begin{equation}
E_r= \sum^{N}_{i=1}[\frac{d_{r}(i)}{128 \pi^4 a^2}]|\frac{F}{ \eta_{i}M^{2}}|^2\frac{2}{3}h(x_{i}) \mathcal{I}
\end{equation}
and $\mathcal{I}$ is an integral that depends on the number of lattice sites 
\begin{equation}
\mathcal{I}=\int^{\infty}_{0} d(ap) (ap) \prod_{j=1}^{N-1}\frac{1}{(ap)^{2}+4\sin^2{\frac{j\pi}{2N}}}
\prod_{i=1}^{N-1}\frac{1}{(ap)^{2}+4\sin^2{\frac{i\pi}{2N}}}.
\end{equation}
The function $\mathcal{I}$ behaves like $\mathcal{I} \sim c/N^4$ 
\begin{center}
\begin{tabular}{c r @{.} l c  c}
$N$ &
\multicolumn{2}{c}{$\mathcal{I}$} & $1/N^4$ & c\\
\hline
$2$ & 0& 25 & 0.0625 & 4\\
$3$ & 0&029 & 0.012 &   2.4      \\ 
$3$ & 0&0082   &   0.0039 &    2.1     \\
$5$ & 0 &0032     &0.0016&      2    \\
$6$ & 0 &0015      &0.00077&     2   \\
$7$ & 0 &00079    &0.00042 &   1.7      \\
\end{tabular}
\end{center}
Rescaling $g^2/N= g^2_{4d}$  and taking $\ell=Na$ we find that the sfermion mass scales as
\be 
m^2_{\tilde{f}}\sim  \frac{g^4_{4d}}{(M\ell)^2}\frac{F^2}{M^2}
\ee
which reproduces the results of the Mirabelli-Peskin model \cite{Mirabelli:1997aj,McGarrie:2010kh}.

\subsection{Casimir energy}
The Casimir energy of the lattice can be extracted from the vacuum energy by taking the difference between the lattice propagator and its continuum counterpart.  This will cancel the divergent parts of the momentum integral  in the vacuum energy.  The final answer will be an approximate result that approaches the continuum Casimir energy when the number of lattice sites is infinite.  We start with \refe{VAC}.  The sum of $C$ terms is still given by \refe{cterms}.  We then must solve the UV divergent momentum integral in \refe{VAC}:
\be 
\sum_{k=0}^{N-1} f(\frac{k}{N})= \! \!\int\! \frac{d^4p}{ (2\pi)^4} [\sum_{k=0}^{N-1} \frac{2(ap)^2 \cos^2 \frac{k\pi}{2N} }{2^{\delta_{k0}}(ap)^2+4\sin^2 \frac{k \pi}{2N} }].
\ee
Next we take the difference between the lattice and continuum momentum integral as 
\be
\sum_{k=0}^{N-1} f(\frac{k}{N})-N  \int_{0}^{\infty} ds f(s)=  \frac{2}{(4\pi)^2\ell^4} \mathcal{S}(N).
\ee
The divergent parts of the lattice and continuum limit of this function will cancel out, leaving the finite mass dependent parts.  Additional steps may be found in the appendix which follow the procedure of \cite{Bauer:2003mh}.  The renormalized function $\mathcal{S}(N)$ is given by 
\begin{align}
\mathcal{S}(N)=-[N^4 & \sum^{N-1}_{k=1}\cos^2 (\frac{k\pi}{N})(\Delta(k/N))^2\log (\Delta(k/N)) \\&-N^5\!\!\int_{0}^{\infty}\! \!\!ds \cos^2(\frac{s\pi}{2})(\Delta(s))^2\log (\Delta(s))]
\end{align}
where 
\be
\Delta(\frac{k}{N})=(a m(\frac{k}{N}))^2= 4\sin^2 \frac{k\pi}{2N}
\ee
In the continuum limit $\mathcal{S}(N)$ is 
\be
\lim_{N\rightarrow \infty} S(N)\rightarrow 3\zeta(5).
\ee  For the Casimir energy, we obtain
\be 
\mathcal{E}_{\text{Casimir}} =- \sum_{r}\sum^{N}_{i=1} 
\frac{g^2_{ r}}{N} \frac{d_{g} d_{r}(i)}{(4\pi)^4}\frac{2}{3\ell^4}|\frac{F}{ \eta_{i}M^2}|^2h(x_{i}) \mathcal{S}(N).
\ee
This result agrees with the Casimir energy found in the Mirabelli-Peskin model \cite{Mirabelli:1997aj}  when $\lim_{N\rightarrow \infty} S(N)$. The Casimir energy for the periodic case is similarly obtained.
\section{Summary and conclusion}\label{conclusion}
In this paper we combine the framework of encoding generic hidden sector in terms of current correlators \cite{Meade:2008wd}, with the four dimensional construction of supersymmetric extra dimensions on a lattice \cite{Csaki:2001em}.  This extends previous lattice constructions of supersymmetry breaking so that different hidden sectors may be explored.  We have demonstrated that the low energy description of this model matches that of extra dimensional supersymmetry breaking on an interval \cite{McGarrie:2010kh}.  In particular we have shown that when the scale of the lattice is much smaller than the characteristic scale of the hidden sector $\frac{1}{(Na)^2}<<M^2$, then for a perturbative messenger sector sfermion masses are suppressed by an additional factor $\frac{1}{(Na M)^2}$ relative to pure four dimensional gauge mediation. This suppression arises due to the suppression of momenta in the effective one loop diagrams generating sfermion masses.

A recent application of this lattice type construction that may arise naturally in metastable supersymmetry breaking models is identified in \cite{Green:2010ww}.
It would also be interesting to explore how $D$-term supersymmetry breaking \cite{Carpenter:2008rj,Dumitrescu:2010ca} may arise in the lattice picture and to understand how those effects can be mediated to the visible sector lattice site.  An unresolved issue in general gauge mediation is that of the mu problem in the Higgs sector \cite{Komargodski:2008ax}. A lattice construction of a warped slice of $AdS_{5}$ should also match the continuum limit found in \cite{McGarrie:2010yk}.  These topics are open for further research.
\paragraph{Acknowledgments} 
I would like to thank Steven Thomas, Rodolfo Russo, Daniel C. Thompson and William Black for useful discussions.  I am funded by STFC.
\appendix
\section{Regularising and renormalising the Casimir energy}
The integral we need to extract the finite part from is 
\be 
\sum_{k=0}^{N-1} f(\frac{k}{N})= \! \!\int\! \frac{d^4p}{ (2\pi)^4} [\sum_{k=0}^{N-1} \frac{2(ap)^2 \cos^2 \frac{k\pi}{2N} }{2^{\delta_{k0}}(ap)^2+4\sin^2 \frac{k \pi}{2N} }].
\ee
We follow the steps outlined in \cite{Bauer:2003mh}.  This integral is UV divergent.  We would like to extract from it the lattice dependent finite part that determines Casimir energy.  We will subtract from it the continuum limit of this function\footnote{There is a subtlety here associated with the difference in normalisation of the zero mode with respect to the rest of the Kaluza Klein tower. As a result the $s=0$ overcounts the zero mode piece. However as the zero modes are massless they will actually contribute nothing to the regularised answer and mat be ignored.}:
\be
N \int_{0}^{\infty} ds f(s) = N \int_{0}^{\infty} ds  \int\! \frac{d^4p}{ (2\pi)^4} [ \frac{2(ap)^2 \cos^2 \frac{s\pi}{2} }{2^{\delta_{s0}}(ap)^2+4\sin^2 \frac{s \pi}{2} }]
\ee
Using 
\be
 \! \!\int\! \frac{d^4y}{ (2\pi)^4}\frac{y^2}{(y^2+\Delta)^\alpha}= \frac{1}{(4\pi)^{d/2}}\frac{d}{2}\frac{\Gamma(\alpha-d/2-1)}{\Gamma(\alpha)}(\Delta)^{1+d/2-\alpha},
\ee
we set $\alpha=1$ and use $d=4-2\epsilon$ to obtain
 \be
\frac{2N}{a^4} \int_{0}^{\infty}  \! \!ds \cos^2 (\frac{s \pi }{2}) \frac{\Delta^2}{(4\pi)^{2}}(2-\epsilon)\Gamma(\epsilon-2)e^{-\epsilon \log(4\pi)-\epsilon \log(\Delta)}.
\ee
We may use
\be
\Gamma(\epsilon-2) =\frac{1}{2\epsilon}+(\frac{3}{4}-\frac{\gamma}{2})+O(\epsilon)
\ee
to give
\be
N \int_{0}^{\infty} ds f(s) = \frac{2N}{a^4} \int_{0}^{\infty} \! \! ds \cos^2 (\frac{s \pi }{2}) \frac{\Delta^2}{(4\pi)^{2}}[\frac{1}{\epsilon}+ (1-\gamma)-\log (4\pi)-\log (\Delta)+ O(\epsilon)]
\ee
Defining 
\be
\sum_{k=0}^{N-1} f(\frac{k}{N})-N  \int_{0}^{\infty} ds f(s)=  \frac{2}{(4\pi)^2\ell^4} \mathcal{S}(N)
\ee
\begin{align}
\mathcal{S}(N)=-[N^4 & \sum^{N-1}_{k=1}\cos^2 (\frac{k\pi}{N})(\Delta(k/N))^2\log (\Delta(k/N)) \\&-N^5\!\!\int_{0}^{\infty}\! \!\!ds \cos^2(\frac{s\pi}{2})(\Delta(s))^2\log (\Delta(s))]
\end{align}
where
\be
\Delta(s)=(a m(s))^2= 4\sin^2 \frac{s \pi}{2}
\ee
In the limit that $ N\rightarrow \infty$ , the mass eigenstates will return to that of a contiuum $S^1/\mathbb{Z}_{2}$ namely $m_{k}=\frac{n\pi}{\ell}$.  We use the Abel-Plana formula \cite{Saharian:2000xx,Bauer:2003mh}
\be
\sum_{k=0}^{N-1} f(\frac{k}{N})-N  \int_{0}^{\infty} ds f(s)=  \frac{1}{2}f(0)+ i\int_{0}^{\infty}dn \frac{f(+in)-f(-in)}{\exp(2\pi n)-1}
\ee
to extract the continuum limit of the Casimir energy:
\be
\lim_{N\rightarrow \infty} S(N)\rightarrow \int \frac{d^4y}{(4\pi)^2} \frac{y}{e^y-1}=3\zeta(5).
\ee

\providecommand{\href}[2]{#2}\begingroup\raggedright\endgroup

\end{document}